\documentclass[twocolumn,nonshowpacs,preprintnumbers,amsmath,amssymb]{revtex4}

\usepackage{amsmath, mathtools, verbatim}
\newcommand{\ra}{\rightarrow}
\newcommand{\sR}{\mathbb R}
\newcommand{\second}{{\prime\prime}}
\newtheorem{theorem}{Theorem}

\begin{document}
\title{Comments on\\
"Nonextensive Entropies derived from \\
Form Invariance of Pseudoadditivity"}

\author{Velimir M. Ili\'c$^1$}\email{velimir.ilic@gmail.com}
\author{Miomir S. Stankovi\'c$^2$}\email{miomir.stankovic@gmail.com}

\affiliation{$^1$ Mathematical Institute of the Serbian Academy of Sciences and Arts, Kneza Mihaila 36, 11000 Beograd, Serbia}
\affiliation{$^2$ University of Ni\v s, Faculty of Occupational Safety, \v Carnojevi\'ca 10a, 18000 Ni\v s, Serbia}

\date{\today\\ \bigskip\bigskip\bigskip\bigskip\bigskip}

\begin{abstract}

Recently, Suyari has defined nonextensive information content
measure with unique class of functions which satisfies certain set
of axioms. Nonextensive entropy is then defined as the appropriate
expectation value of nonextensive information content [H. Suyari,
Phys. Rev E 65 066118 (2002)]. In this comment we show that the
class of functions determined by Suyari's axioms is actually wider
than the one given by Suyari and we determine the class.
Particularly, an information content corresponding to
Havrda-Charv\'at entropy satisfies Suyari's axioms and does not
belong to the class given by Suyari but belongs to our class.
Moreover, some of the conditions from Suyari's set of axioms are
redundant, and some of them can be replaced with more intuitive
weaker ones. We give a modification of Suyari's axiomatic system
with these weaker assumptions and define the corresponding
information content measure.

\end{abstract}
\pacs{02.50.-r  05.20.-y  89.70.+c} \maketitle

\newcommand{\limes}[2]{\mathop {\lim }\limits_{#1 \to #2}}
\newcommand{\qlim}[1]{\limes{q}{#1}}
\newcommand{\Z}{\sum\limits_{j=1}^n {p_j^{{\varphi(q)} +1}}}
\newcommand{\newfactor}{{\alpha(q)}}

\section{Introduction}

In \cite{Suyari_02}, Suyari gives a nonextensive generalization of
axioms for standard
information content \cite{Viterbi_79}. 
Nonextensive information content is defined as a function $I_q(p)$
of two variables $q\in \sR^+$ and $p \in \left( {0,1} \right]$,
which satisfies the following axioms.


\begin{description}
\item{[T0]} $I_1(p) = \qlim{1}I_q \left( p \right)=- k \ln p$, $k \neq
0$;
\item{[T1]} $I_q$ is differentiable
with respect to $p\in\left( {0,1}\right)$ and $q\in \sR^+$,
\item{[T2]} $I_q\left( {p} \right]$
is convex with respect to $p\in \left(0,1\right]$ for any fixed
$q\in \sR^+$;
\item{[T3]} there exists a function $\varphi: \sR^+ \to \sR$ such that
\begin{equation}
{I_q\left( {p_1p_2} \right)\over{k}}={I_q\left(
{p_1}\right)\over{k}} +{I_q\left( {p_2} \right)\over{k}}
+\varphi\left( q \right)\cdot{I_q\left( {p_1}\right)\over{k}}\cdot
{I_q\left( {p_2} \right)\over{k}} \label{pseudoadditivity}
\end{equation}
for any $p_1,p_2\in \left( {0,1} \right]$, where $\varphi \left( q
\right)$ is differentiable with respect to $q\in \sR^+$ and
\begin{equation}
\qlim{1}{{d\varphi \left( q \right)} \over {dq}}\ne 0,\;\;
\mathop {\lim }\limits_{q\to 1}\varphi \left( q \right)=\varphi
\left( 1 \right)=0, \;\;
\varphi \left( q \right)\ne
0\; \left( {q\ne 1} \right). \label{boundary}
\end{equation}
\end{description}
For $q=1$ axioms [T1]$\sim$[T3] reduce to axioms of standard
information content \cite{Viterbi_79}, which is given with [T0].

Suyari claims that nonextensive information content $I_q\left( p
\right)$ obtained from the axioms [T0]$\sim$[T3] is uniquely
determined with the function
\begin{equation}
\label{information content s}
I_q\left( p \right) =k\cdot{{p^{-
\varphi \left( q \right)}-1} \over {\varphi \left( q\right)}}
\end{equation}
for $q \neq 1$, where $k$ is a positive constant and
\begin{equation}
\label{introduction: phi condition} \varphi \left( q \right)+1\ge
0\quad {\hbox{for any}}\;{q\in \sR^+}.
\end{equation}






However, the class of functions which satisfy [T0]$\sim$[T3] is
actually wider than the class given by Suyari. For example, the
information content given by
\begin{equation}
\label{information content example}
I_q\left( p \right)=%
- k \ln 2 \cdot {p^{1-q} - 1 \over 2^{1-q} - 1}.
\end{equation}
clearly does not belong to the class determined by equation
(\ref{information content s}). On the other hand,
(\ref{information content example}) obviously satisfies
[T1]$\sim$[T3]. Moreover, if we introduce $\gamma_p(q) = p^{1-q} -
1$ and $\delta(q) = 2^{1-q} - 1$, and divide the numerator and
denominator with $1-q$, we get
\begin{equation}
I_q\left( p \right)=%
- k  \ln p \cdot {\ln(1 + \delta(q))^{1 / \delta(q)} \over \ln(1 +
\gamma_p(q))^{1 / \gamma_p(q)}}.
\end{equation}
Now, [T0] straightforwardly follows if we take the limit, noting
that $\gamma_p(q) \rightarrow 0$ and $\delta(q) \rightarrow 0$ as
$q \rightarrow 1$ and keeping in mind that $\mathop {\lim
}\limits_{\alpha \to 0} (1 + \alpha)^{1 \over \alpha} = e$.  
In section \ref{axioms: correction} we determine the class to
which both information contents, (\ref{information content s}) and
(\ref{information content example}), belong. After that, in
section \ref{entropy}, we show that averaging of information
content (\ref{information content example}) yields the
Havrda-Charv\'at entropy \cite{Havrda_Charvat_67}.

In section \ref{axioms: new} we consider the redundancy of some
conditions from [T0]$\sim$[T3].
%
%
Particularly, the conditions $ \qlim{1}\varphi \left( q
\right)=\varphi \left( 1 \right)=0, \;\;\hbox{and}\;\; \varphi
\left( q \right)\ne 0\; \left( {q\ne 1} \right)$ from [T3] are
redundant, since they follow from [T0]$\sim$[T2].
%
Moreover, the condition $\qlim{1}{{d\varphi \left( q \right)}
\over {dq}}\ne 0$ from [T3] is introduced to ensure that
L'Hospital's rule can be applied on (\ref{information content s})
to prove that [T0] holds. However, this is unnecessary, since the
satisfaction of [T0] can be shown without the use of L'Hospital's
rule in elementary way using equality $\mathop {\lim
}\limits_{\alpha \to 0} (1 + \alpha)^{1 \over \alpha} = e$, as we
have shown in the previous paragraph. 
In addition, in section \ref{axioms: new} we give an alternative
proof for the information content without use of the
differentiability condition from [T1] so that the
differentiability condition can be weakened to continuity.


\section{New class of information context functions - correction to Suyari's
proof}
\label{axioms: correction}

\begin{theorem}\rm
\label{axioms: correction: suyari solution} Let $I_q(p)$ be a
function of two variables $q\in \sR^+$ and $p \in \left( {0,1}
\right]$, which satisfies axioms [T0]$\sim$[T3]. Then, $I_q(p)$ is
uniquely determined with

%
\begin{equation}
\label{axioms: correction: information content}
I_q\left( p \right)=%
{k \over {\varphi \left( q \right)}}%
\cdot \left( p^{\alpha(q)} -1 \right),
\end{equation}
where $k \in \sR^+$ and
\begin{description}
\item{(a)} $\alpha \left( q \right)$ is differentiable with
respect to any $q\in \sR^+$ and
\begin{equation}
\label{axioms: correction: limes alpha condition} \qlim{1} {\alpha
\left( q \right) \over \varphi \left( q \right)} = -1;
\end{equation}

\item{(b)} it holds that
\begin{equation}
\label{axioms: correction: (b) property} \alpha(q) \in
\begin{dcases}
(-\infty,0] 
\;\; &\hbox{for} \;\; \varphi(q) > 0 \\
\quad
[0 , 1] \;\; &\hbox{for} \;\; \varphi(q) < 0.
\end{dcases}
\end{equation}

\end{description}

\end{theorem}

\textbf{Remark 1.}
Positivity of constant $k$ directly follows from the convexity of
$I_1(p)$. Positivity of $k$ further implies nonnegativity of
information content (\ref{axioms: correction: information
content}).

\textbf{Remark 2.}
For the case $\alpha(q)=-\varphi(q)$ information content
(\ref{axioms: correction: information content}) reduces Suyari's
information content (\ref{information content s}) and the
condition (\ref{axioms: correction: (b) property}) reduces to
(\ref{introduction: phi condition}).

\vskip 0.5cm

\textbf{Proof.}
Following Suyari \cite{Suyari_02}, we substitute $p_1=p$ and
$p_2=1+\Delta$ in equation (\ref{pseudoadditivity}) and divide the
equation with $\Delta p$. By taking the limit $\Delta \to 0$ on
both sides of the resulting equation and using differentiability of
$I_q(p)$, we obtain
%
%
%
\begin{equation}
\label{differntial0} {{dI_q\left( p \right)} \over {dp}}={\beta(q)
\over{k}}\cdot {{k+\varphi \left( q \right)I_q\left( p \right)}
\over p},
\end{equation}
where the function $\beta(q)$ is defined with $\beta(q)
\equiv\mathop {\lim }\limits_{\Delta\to 0-}{{I_q\left( {1+\Delta }
\right)} \over \Delta }$.
This can be solved analytically for every $q$; the rigorous
solution is
\begin{equation}
\label{Iq solution beta} I_q\left( p \right)=k\cdot{{\left( {C(q)
p^{\beta(q)} } \right)^{{\varphi \left( q \right)}\over{k}}-1}
\over {\varphi \left( q \right)}},
\end{equation}
where $C(q)$ and $\beta(q)$ are functions of $q$ \footnote{At this
point, Suyari implicitly makes assumption that $\beta(q)=\beta$
and $C(q)=C$ are constant functions. However, it can be easily
shown that partial differential equation (\ref{differntial0})
holds for nonconstant functions, too.}. If we introduce
\begin{equation}
K(q) =C(q)^{\varphi \left( q \right) \over k}
\quad\mbox{and}\quad \alpha(q) = {\beta(q) \varphi \left( q
\right) \over k },
\end{equation}
expression (\ref{Iq solution beta}) becomes
\begin{equation}
\label{Iq solution} I_q\left( p \right)={k \over \varphi(q)} \cdot
\left({K(q) \cdot p^{\alpha(q)} - 1}\right).
\end{equation}
For $p_1=p_2=1$, [T3] reduces to
\begin{equation}
\label{axioms: Iq(1) condition} {I_q\left( 1\right)\over{k}}
\left(1 + \varphi \left( q \right) {I_q\left( 1\right)\over{k}}
\right)=0,
\end{equation}
with
\begin{equation}
I_q\left( 1 \right)={k \over \varphi(q)} \left( K(q) -1  \right).
\end{equation}

Equality (\ref{axioms: Iq(1) condition}) can be satisfied if and
only if $I_q\left( 1\right) = 0$ or $I_q\left( 1\right) = -
{k\over\varphi\left(q\right)}$. The former case is not allowed
since it is inconsistent with [T0]. Accordingly, $I_q\left(
1\right) = 0$, or equivalently $K(q)=1$,
and (\ref{Iq solution}) reduces to
(\ref{axioms: correction: information content}).

We will now prove the property (a). Differentiability of
$\alpha(q)$ follows from (\ref{axioms: correction: information
content}), since $I_q(p)$ and $\varphi(q)$ are differentiable with
respect to $q$.
%
Moreover, by use of [T3] we have $\qlim{1} \alpha(q) = 0$, and
%
which implies that there exists $a > 0$ such that $\alpha(q) \neq
0$ for $q \in (1 - a, 1 + a) \setminus \{1\}$. Otherwise there
exists a sequence $q_1, q_2, \dots$ converging to $1$, such that
$\alpha(q_i)=0$ and $I(q_i)=0$. This contradicts [T0] since
$I(q_i) \ra 0$ as $i \ra \infty$.
%
%
Let us introduce $\gamma(q)= p^{\alpha(q)} - 1$. Using $\gamma(q)
\ra 0$ when $q \rightarrow 1$ and $(1 + t )^{1 \over t} \ra e$
when $t \rightarrow 0$, we have
\begin{align}
\label{limit Iq}
&\qlim{1} I_q\left( p \right) =%
\qlim{1} {k \over {\varphi \left( q \right)}}%
\cdot{ \left( p^{\alpha(q)} - 1 \right) }=\nonumber\\ %
&=k \cdot \qlim{1} {\alpha(q) \over \varphi(q)} \cdot {
p^{\alpha(q)} - 1  \over \alpha(q)}=\nonumber\\%
&=k \cdot \qlim{1} {\alpha(q) \over \varphi(q)}
\cdot  {\ln p \over \ln \left(1 + \gamma(q)\right)^{1 \over \gamma(q) }}=\nonumber\\%
&=k \ln p \cdot \qlim{1} {\alpha(q) \over \varphi(q)}.
\end{align}
According to [T0], we have $\qlim{1} I_q = - k \ln p$ and property (a) follows.

Property (b) can be proven by taking the second derivative of
${I_q\left( p \right)}$ with respect to $p$, which should be
nonnegative for any fixed $q\in \sR^+$, since $I_q(p)$ is convex
by [T2]. Thus, we can derive a constraint
\begin{equation}
\label{alpha constraint general}%
k \cdot {\alpha(q) \over \varphi(q)}\cdot\left(\alpha(q) -1
\right)\ge 0
\end{equation}
for any $q\in \sR^+$. Since $k$ is positive, the
constraint (\ref{alpha constraint general}) is satisfied if
\begin{align}
\label{alpha constraint cases general}
\alpha(q) \in
\begin{dcases}
(-\infty,0] \cup [1, \infty) \;\; &\hbox{for} \;\; \varphi(q) > 0 \\
\quad\quad\quad [0 , 1] \;\; &\hbox{for} \;\; \varphi(q) < 0.
\end{dcases}
\end{align}
We will now prove that $\alpha(q) \not\in [1, \infty)$ for $\varphi(q) > 0$, by
contradiction.

Recall that $\varphi(q) \neq 0$ for $q \neq 1$, by [T3] and according
to the intermediate value theorem,  $\varphi(q)$ does not change sign on $(0,1)$ nor on $(1,\infty]$.
Let
\begin{equation}
\label{axioms: correction: varphi(q)>0}
\varphi(q)>0, \quad \mbox{for all} \quad q>1,
\end{equation}
%
%
%
and let $\alpha(q) \in [1, \infty)$ for some $q > 1$ (the case $0 < q <1$ can be considered analog).

%
%
According to (\ref{axioms: correction: limes
alpha condition}), for arbitrary small $\varepsilon>0$ there
exists $\delta > 0$ such that $|\alpha(q^\prime)| < \varepsilon$
for $q^\prime \in (1, 1+ \delta)$. Hence, $\alpha(q^\prime) <
\varepsilon$ and $\alpha(q)\geq 1$ for some $q,q^\prime>1$. According
to the intermediate value theorem, if $\mu \in (\varepsilon, 1)$,
then $\alpha(q^\second)=\mu$ for some $q^{\second}>1$. If we
combine the condition $\alpha(q^\second) \in (\varepsilon, 1)$ for
some $q^\second > 1$, with condition (\ref{alpha constraint cases
general}), we get
\begin{equation}
\label{axioms: correction: varphi(q)<0}
\varphi(q^\second) < 0, \quad \mbox{for some} \quad q^\second>1.
\end{equation}
According to the intermediate theorem, (\ref{axioms: correction:
varphi(q)>0}) and (\ref{axioms: correction: varphi(q)<0}) imply
that there exists $q^\prime > 1$ such that $\varphi(q^\prime) = 0$
for some $q^\prime > 1$. This contradicts the condition  $\varphi
\left( q \right)\ne
0\; \left( {q\ne 1} \right)$ from [T3] and proves property (b).
%

%
%

\section{Havrda-Charv\'at entropy as expected information content}
\label{entropy}

Nonextensive entropy of distribution $p$, $S_q\left( p \right)$,
is defined as the appropriate expectation value of 
$I_q\left( p \right)$,
\begin{equation}
\label{gexp: nonextensive entropy}%
S_q\left(p \right)\equiv %
E_{q,p}\left[ {I_q\left( p \right)} \right].
\end{equation}
Similarly, the Kullback-Leibler ($KL$) divergence between
distributions $p^A$ and $p^B$ is defined by means of the
information contents difference,
\begin{equation}
K_q\left( {p^A\;\left\| {\;p^B} \right.} \right) =E_{q,p^A}\left[
{I_q\left( {p^B} \right)-I_q\left( {p^A} \right)} \right].
\label{KLentropy1}
\end{equation}

%


According to Suyari, the expectation should be chosen so that the $KL$ divergence is nonnegative \cite{Suyari_02}. 
%
One possible choice is 
generalized $q$-expectation
\begin{equation}
\label{entropy: exp org suyari} E_{q,p}^{\hbox{g-org}}\left[ X
\right]\equiv\sum\limits_{i=1}^W {p_i^{- \newfactor + 1}}X_i.
\end{equation}
By use of Jensen's inequality following Suyari's proof
(\cite{Suyari_02}), it can be straightforwardly shown that
generalized expectations (\ref{entropy: exp org suyari}) can be
combined with information content (\ref{axioms: correction:
information content}) leading to nonnegative Kullback-Leibler
divergence.

Generalized nonextensive entropy (\ref{gexp: nonextensive
entropy}) is now defined by taking the expectation (\ref{entropy:
exp org suyari}) of information content (\ref{axioms: correction:
information content}):
\begin{align}
\label{invar: S unnor}%
S_q^{\hbox{g-org}}\left( p \right) &\equiv%
E_{q,p}^{\hbox{g-org}}\left[ {I_q\left( p \right)} \right]= \nonumber\\ %
&={k \over \varphi(q)} \cdot%
\left[ {1-\sum\limits_{i=1}^W {p_i^{-\newfactor +1}}}\right].
\end{align}

Let us now define
\begin{equation}
\varphi(q)={1 - 2^{1-q} \over \ln 2} \quad \text{and}\quad
\alpha(q) = 1 -q.
\end{equation}
Functions $\varphi(q)$ and $\alpha(q)$ satisfy conditions from
Theorem \ref{theo: information content}, which means that
information content from introductional example (\ref{information
content example})
belongs to the class determined by (\ref{information content}).
Moreover, if we choose $k = 1 / \ln 2$, generalized entropy
(\ref{invar: S unnor}) reduces to
\begin{equation}
S_q^{\hbox{hc-org}}\left( p \right) =%
{1 - \sum\limits_{i=1}^W {p_i^q } \over 1 - 2^{1-q}},
\end{equation}
%
which represents the Havrda-Charv\'at entropy
\cite{Havrda_Charvat_67}.

\newcommand{\Ra}{\Rightarrow}

\section{New axiomatic system}
\label{axioms: new}

As we noted in introduction, some conditions from [T0]$\sim$[T3]
are redundant and can be omitted, and some of them can be
weakened so that [T0]$\sim$[T3] still lead to information content
with form (\ref{axioms: correction: information content}).

First, the condition $I_1(p) = - k \ln p$ from [T0] can be
substituted in (\ref{pseudoadditivity}) for $q=1$ and we get
$\varphi(1)=0$. Accordingly, condition $\varphi(1)=0$ in [T3] is
redundant. Hence, we suggest:

\begin{description}

\item{1.} Relax the assumption $\varphi(1) = 0$ from [T3].

\end{description}

%
%

Axiom [T1] requires differentiability of $I_q(p)$ with respect to
$p$. This condition is necessary when differential equation
(\ref{differntial0}) is constructed in order to get the form
(\ref{axioms: correction: information content}).
However, the functional equation (\ref{pseudoadditivity}) can be
solved without using the condition that $I_q(p)$ is differentiable
with respect to $p$. If we introduce the transformation
\begin{equation}
\label{new: f substitution}%
I_q(p) ={k \over \varphi(q)} \left( f_q(p) - 1\right),
\end{equation}
equation (\ref{pseudoadditivity}) becomes Cauchy's
functional equation, $f_q(p_1 p_2) = f_q(p_1) f_q(p_2)$, 
which has unique continuous solution \cite{Aczel_06}
\begin{equation}
\label{new: cauchhy fun eq solution} f_q(p) = p^{\alpha(q)},
\end{equation}
where $\alpha(q)$ is arbitrary real function of $q$. By
substituting (\ref{new: cauchhy fun eq solution}) in (\ref{new: f
substitution}), we obtain  (\ref{axioms: correction: information content}). In this way, the
solution is obtained only using the assumption that $I_q(p)$ is
continuous with respect to $p$. Therefore, we suggest:

\begin{description}
\item{2.}
In [T1], replace the assumption ``$I_q(p)$ is differentiable with respect to $p$"  with ``$I_q(p)$ is continuous with respect to
$p$".
\end{description}

In Suyari's proof the assumption about differentiability of
$\varphi(q)$ and the condition $\qlim{1}{{d\varphi \left( q
\right)} \over {dq}}\ne 0$ from [T3] are used when L'Hospital's
rule is applied for taking the limit of (\ref{Iq solution}).
However, it can be done in an elementary manner by use of
continuity and equality $\mathop {\lim }\limits_{\alpha \to 0} (1
+ \alpha)^{1 \over \alpha} = e$, as we did in (\ref{limit Iq}).
Therefore, we make the following suggestions:

\begin{description}
\item{3.}
Relax the assumption $\qlim{1}{{d\varphi \left( q
\right)} \over {dq}}\ne 0$ from [T3].
\end{description}

Axioms [T1] and [T3] require that $I_q(p)$ and $\varphi(q)$ be
differentiable with respect to $q$. These conditions are used only
to show that $\alpha(q)$ is differentiable as stated in property
(a); and differentiability of $\alpha(q)$ is then used only to
conclude that $\alpha(q)$ is continuous, which, in turn, is used to prove
property (b). Therefore, we make the following suggestion:

\begin{description}

\item{4.}
Relax the assumption about differentiability of
$I_q(p)$ and $\varphi(q)$ with respect to $q$, by supposing only that $I_q(p)$ and $\varphi(q)$
are continuous with respect to $q$.

\end{description}
Note that if these assumptions are accepted, the statement
``$\alpha(q)$ is differentiable" from property (a) should be
changed to ``$\alpha(q)$ is continuous".

%
%
%
%
%
%
%
%

The discussion led us to the following modification of Theorem
\ref{axioms: correction: suyari solution}.


%

\begin{theorem}\rm
\label{theo: information content}
Let $I_q(p)$ be a function of two variables $q\in \sR^+$ and $p
\in \left( {0,1} \right]$, which satisfies the following axioms:

\begin{description}
\item{[T0]} $I_1\left( p \right)=- k \ln p$, $k \neq 0$
\item{[T1]} $I_q$ is continuous with respect to $p\in\left( {0,1}\right]$ and $q\in \sR^+$,
\item{[T2]} $I_q\left( {p} \right)$ is convex with respect to $p\in \left(0,1\right]$ for any fixed
$q\in \sR^+$,
\item{[T3]} there exists a function $\varphi :R\to R$ such that
\begin{equation}
{I_q\left( {p_1p_2} \right)\over{k}}={I_q\left(
{p_1}\right)\over{k}} +{I_q\left( {p_2} \right)\over{k}}
+\varphi\left( q \right)\cdot{I_q\left( {p_1}\right)\over{k}}\cdot
{I_q\left( {p_2} \right)\over{k}} 
\end{equation}
for any $p_1,p_2\in \left( {0,1} \right]$, $\varphi(q) \neq 0$ for
$q \neq 1$ and $\varphi(q)$ is continuous.
\end{description}
Then,
\begin{equation}
\label{information content}
I_q\left( p \right)=%
{k \over {\varphi \left( q \right)}}%
\cdot \left( p^{\alpha(q)} -1 \right),
\end{equation}
where $k$ is a positive constant and
\begin{description}

\item{(a)} $\alpha \left( q \right)$ is continuous with
respect to any $q\in \sR^+$, and
\begin{equation}
\qlim{1} {\alpha \left( q \right) \over \varphi \left( q \right)}
= -1;
\end{equation}

\item{(b)} it holds that
\begin{equation}
\alpha(q) \in
\begin{dcases}
(-\infty,0] 
\;\; &\hbox{for} \;\; \varphi(q) > 0 \\
\quad
[0 , 1] \;\; &\hbox{for} \;\; \varphi(q) < 0.
\end{dcases}
\end{equation}

\end{description}

\end{theorem}
%
%
%
%


\begin{acknowledgments}
This research was supported by the Ministry of Science and
Technological Development, Republic of Serbia, Grant No. III44006
\end{acknowledgments}

\bibliographystyle{plain}
\bibliography{Comments_Suyari}

\begin{thebibliography}{1}

\bibitem{Aczel_06}
J.~Aczel.
\newblock {\em Lectures on Functional Equations and Their Applications}.
\newblock Dover Books on Mathematics. Dover Publications, 2006.

\bibitem{Havrda_Charvat_67}
Jan Havrda, František Charv\'at, Jan Havrda, and Frantisek Charvât.
\newblock Quantification method of classification processes: Concept of
  structural a-entropy.
\newblock {\em Kybernetika}, 1967.

\bibitem{Suyari_02}
Hiroki Suyari.
\newblock Nonextensive entropies derived from form invariance of
  pseudoadditivity.
\newblock {\em Phys. Rev. E}, 65:066118, Jun 2002.

\bibitem{Viterbi_79}
Andrew~J. Viterbi and James~K. Omura.
\newblock {\em Principles of Digital Communication and Coding}.
\newblock McGraw-Hill, Inc., New York, NY, USA, 1st edition, 1979.

\end{thebibliography}

\end{document}